\documentclass[fleqn,twoside,twocolumn,nofootinbib]{revtex4} 
\usepackage[sec]{ujp} 
\begin{document}
\title[SPIN-1/2 ASYMMETRIC DIAMOND ISING--HEISENBERG CHAIN]
{SPIN-1/2 ASYMMETRIC DIAMOND ISING--HEISENBERG CHAIN}%
\author{B.М.~Lisnii}
\affiliation{Institute for Condensed Matter Physics, Nat. Acad. of Sci. of Ukraine}
\address{1, Sventsits’kyi Str., Lviv 79011, Ukraine}
\email{lisnyj@icmp.lviv.ua}
\udk{538.953, 538.955} \pacs{75.10.Pq, 75.40.Cx,\\[-3pt] 75.10.Jm}
\razd{\secix}

\setcounter{page}{1237}%
\maketitle

\begin{abstract}
The ground state and the thermodynamics of a spin-1/2 asymmetric
diamond Ising--Heisenberg chain are considered. For the $XYZ$
anisotropic Heisenberg interaction, the exact calculations of the
free energy, entropy, heat capacity, magnetization, and magnetic
susceptibility are performed using the method of
decoration-iteration transformation. In the case of
antiferromagnetic interactions (Ising and $XXZ$ anisotropic
Heisenberg ones), the ground state, magnetization process,
temperature dependence of the magnetization, magnetic
susceptibility, and heat capacity are investigated. The influence of
geometric frustration and quantum fluctuations on these
characteristics is studied.
\end{abstract}

\section{Introduction}

One of the interesting objects of statistical physics is exactly
solvable decorated chains, whose structure is formed by the
decoration of a primitive cell of a spin-1/2 Ising chain. The exact
solution of these decorated chains is obtained using the method of
decoration-iteration transformation \cite{s5,s6}. They particularly
include the following one-dimensional models: spin-(1/2, $S > 1/2$)
\cite{ka97}, ferromagnetic-ferromagnetic-antiferromagnetic
\cite{oh03}, and diamond \cite{valpa08} Ising chains, simple
\cite{dos1,s3,lis1}, diamond \cite{dos2,scmp09}, tetrahedral
\cite{valjpcm08}, and sawtooth \cite{ohcmp09} Ising--Heisenberg
chains, Ising--Heisenberg chain with triangular Heisenberg
plaquettes \cite{ohprb09}, and asymmetric diamond Ising--Hubbard
chain \cite{dos3,lis2}. Decorated chains are convenient objects for
studying the appearance of intermediate plateaus on the
magnetization curve and additional maxima on the temperature
dependence of the heat capacity, as well as the mutual influence of
the geometrical frustration and quantum fluctuations. All these
phenomena take place in real systems \cite{s3,ki05l,ki05ptp}.

We consider a spin-1/2 asymmetric diamond Ising--Heisenberg chain
with the $XYZ$ anisotropic Heisenberg interaction between decoration
spins. It has the same diamond structure as that considered in
\cite{dos2} and the same asymmetry of the Ising interaction on bonds
along the diamond sides as in \cite{dos3}. The decoration-iteration
transformation \cite{s5,s6} is used to exactly calculate the
thermodynamic characteristics of this chain. For the
antiferromagnetic Ising interaction and the antiferromagnetic $XXZ$
Heisenberg interaction, at which the system is geometrically
frustrated, we consider the ground state, magnetization process,
temperature dependence of the magnetization, magnetic
susceptibility, and heat capacity. The effect of the Ising and
Heisenberg interactions on these characteristics is studied.

\section{Exact Solution of the Model}

\begin{figure}
\includegraphics[width=\column]{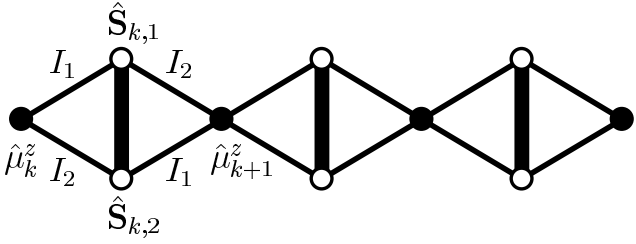}
\caption{Fragment of an asymmetric diamond Ising--Heisenberg
chain. Spins $\hat\mu^z_{k}$, $\hat\mu^z_{k+1}$ at nodes and spins
$\hat {\bf S}_{k,1}$, $\hat {\bf S}_{k,2}$ at interstitial positions
of the $k$-th primitive cell are marked. The Ising interaction
parameters on bonds along the diamond sides ($I_1$ and $I_2$) are
different for samples with different orientations  \label{fig1}}
\end{figure}

Consider a spin-1/2 asymmetric diamond Ising--Heisenberg chain in
the magnetic field. A primitive cell of this chain is determined by
nodes $k$ and $k{+}1$ (Fig.~\ref{fig1}). They are occupied by
the so-called Ising spins ($\hat\mu^z_{k}$) coupled with neighbors by
the Ising interaction. Two interstitial positions $(k,1)$ and
$(k,2)$ in the primitive cell (Fig.~\ref{fig1}) are occupied by
the so-called Heisenberg spins ($\hat {\bf S}_{k,1}$ and $\hat {\bf
S}_{k,2}$) with the Heisenberg interaction between them. We write the
Hamiltonian of the chain $\hat {\cal H}$ as the sum of
the cell Hamiltonians $\hat{\cal H}_k$:
\[
\hat {\cal H} = \sum\limits_{k=1}^N \hat {\cal H}_k,
\]
\[
\hat {\cal H}_k = J_1 \hat S^x_{k,1}\hat S^x_{k,2} + J_2 \hat
S^y_{k,1}\hat S^y_{k,2} + J_3 \hat S^z_{k,1}\hat S^z_{k,2} +
\]
\[
{} +  \hat\mu^z_{k}(I_1 \hat S^z_{k,1} + I_2 \hat S^z_{k,2}) + \hat
\mu^z_{k+1}(I_2 \hat S^z_{k,1} + I_1 \hat S^z_{k,2}) -
\]
\begin{equation}
{} - \frac{h_{\textrm{I}}}{2} (\hat \mu^z_{k} + \hat \mu^z_{k+1}) -
h_{\textrm{H}} (\hat S^z_{k,1} + \hat S^z_{k,2}), \label{Hk}
\end{equation}
where $N$ is the number of primitive cells; $\hat \mu^z_{k}$ and $\hat
S^\alpha_{k,i}$ ($\alpha = x, y, z$; $i= 1, 2$) are the components
of spin-1/2 operators; $J_1$, $J_2$, and $J_3$ are the parameters of
the Heisenberg interaction; $I_1$, $I_2$ are the parameters of the
Ising interaction on the bonds along the diamond sides
(Fig.~\ref{fig1}); $h_{\textrm{I}}$ and $h_{\textrm{H}}$ are the
magnetic fields acting on the Ising and Heisenberg spins,
respectively. It is worth noting that Hamiltonian (\ref{Hk}) also
corresponds to a simple Ising--Heisenberg chain, in which the Ising
spin interacts with the first ($I_1$) and second ($I_2$) neighbors.
In the particular cases $I_1 = I_2$ and $I_2=0$ (or $I_1=0$),
Hamiltonian (\ref{Hk}) corresponds to the above-considered
Ising--Heisenberg chains: diamond \cite{dos2} and simple
\cite{dos1,lis1} ones.

Let us find the statistical sum of the system ${\cal
Z}=\mbox{Tr}~e^{-\beta \hat {\cal H}}$, where
$\beta=1/k_{\textrm{B}}T$, $k_{\textrm{B}}$ is the Boltzmann
constant, and $T$ is the absolute temperature. With regard for the
commutativity of the Hamiltonians $\hat{\cal H}_k$, the statistical
sum ${\cal Z}$ is partially factorized:
\begin{equation}
{\cal Z} = \mbox{Tr}_{\{\hat\mu^z\}} \prod_{k=1}^N \mbox{Tr}_{\hat
{\bf S}_{k,1},\hat {\bf S}_{k,2}} \exp (-\beta \hat {\cal H}_k),
\label{Z1}
\end{equation}
where $\mbox{Tr}_{\{\hat\mu^z\}}$ means the trace over the states of
all Ising spins and $\mbox{Tr}_{\hat {\bf S}_{k,1},\hat {\bf
S}_{k,2}}$ is the trace over the states of two Heisenberg spins from
the $k$-th cell. Now, we calculate the factor
\[
{\cal Z}_k(\hat\mu_{k}^z, \hat\mu_{k+1}^z) =
\mbox{Tr}_{\hat {\bf S}_{k,1},\hat {\bf S}_{k,2}} \exp (-\beta \hat {\cal H}_k).
\]

For this purpose, we should pass to the matrix representation of the
Hamiltonian $\hat {\cal H}_k$ in the basis constructed by the
eigenstates of the operator ${\hat S}^z_{k,1} {\hat S}^z_{k,2}$:
\[
|\uparrow(\downarrow), \uparrow(\downarrow) \rangle_{k,1;k,2}=
|\uparrow(\downarrow) \rangle_{k,1}|\uparrow(\downarrow) \rangle_{k,2},
\]
where $|\uparrow\rangle_{k,i}$ and $|\downarrow\rangle_{k,i}$ denote the eigenstates of ${\hat S}^z_{k,i}$.
The eigenvalues of the matrix $\hat {\cal H}_k$ have the form
\[
{\cal E}_{1,2} (\hat\mu_{k}^z, \hat\mu_{k+1}^z)= \frac{J_3}{4} -
\frac{h_{\textrm{I}}}{2}(\hat\mu_{k}^z + \hat\mu_{k+1}^z) \pm
\]
\[
\pm \sqrt{ \frac{(J_1 - J_2)^2}{16} + \left (\frac{I_1 +
I_2}{2}(\hat\mu_{k}^z + \hat\mu_{k+1}^z)- h_{\textrm{H}} \right )^2
},
\]
\[
{\cal E}_{3,4} (\hat\mu_{k}^z, \hat\mu_{k+1}^z)= - \frac{J_3}{4} -
\frac{h_{\textrm{I}}}{2}(\hat\mu_{k}^z + \hat\mu_{k+1}^z) \pm
\]
\begin{equation}
\pm \sqrt{ \frac{(J_1 + J_2)^2}{16} + \frac{(I_1 -
I_2)^2}{4}(\hat\mu_{k}^z - \hat\mu_{k+1}^z)^2 }. \label{Ek}
\end{equation}
Having obtained ${\cal Z}_k(\hat\mu_{k}^z, \hat\mu_{k+1}^z) =
\sum\limits_{i=1}^4 e^{-\beta {\cal E}_i (\hat\mu_{k}^z,
\hat\mu_{k+1}^z)}$, we perform the decoration-iteration
transformation \cite{s5,s6}:
\[
{\cal Z}_k(\hat\mu_{k}^z,\hat\mu_{k+1}^z) {=}
A \exp \left(\beta R \hat\mu_{k}^z \hat\mu_{k+1}^z + \beta h_0 (\hat\mu_{k}^z + \hat\mu_{k+1}^z)/2 \right),
\]
where $A$, $R$, and $h_0$ are the transformation parameters determined by the relations
\[
A =\left ({\cal Z}_k(+,+) {\cal Z}_k(-,-) {\cal Z}_k^2(+,-)
\right)^{{1}/{4}},
\]
\[
\beta R = \ln \frac{{\cal Z}_k(+,+) {\cal Z}_k(-,-)}{{\cal Z}_k^{2}(+,-)}, \quad
\beta h_0 = \ln \frac{{\cal Z}_k(+,+)}{{\cal Z}_k(-,-)},
\]
in which the argument ``$\pm$'' means $\pm1/2$. With the use of this
transformation, the calculation of the statistical sum (\ref{Z1}) of the
Ising--Heisenberg chain is reduced to that of the
statistical sum of the spin-1/2 Ising chain with interaction $R$ and
field $h_0$. Applying the known result \cite{s2} to the latter, we
obtain the statistical sum (\ref{Z1}) in the form
\[
{\cal Z} = A^N (\lambda_1^N + \lambda_2^N),
\]
where
\[
\lambda_{1, 2} = e^{\beta R/4} \left( \ch(\beta h_0/2) \pm \sqrt{\sh^2(\beta h_0/2) + e^{-\beta R}} \right) .
\]

After that, the free energy of the cell in the thermodynamic limit takes the form
\[
f = -\frac{1}{\beta} (\ln A + \ln \lambda_1),
\]
which can be used for the calculation of the entropy $s$ and the heat capacity $c$:
\[
s = k_{\textrm{B}} \beta^2  \frac{\partial f}{\partial \beta}~,
\qquad
c = - \beta \frac{\partial s}{\partial \beta}~.
\]
The magnetization of the Ising spins $m_{\textrm{I}}=\langle \hat
\mu_{k}^z + \hat \mu_{k+1}^z \rangle /2$ and their correlation
function $q_{\textrm{I}\textrm{I}}(n)=\langle \hat \mu_{k}^z \hat
\mu_{k+n}^z \rangle$ are the same as for the spin-1/2 Ising chain
with interaction $R$ and field $h_0$ \cite{dos1,dos2,lis1}. That is
why we can use the known results \cite{s2} for them. The
magnetization of the Heisenberg spins $m_{\textrm{H}} = \langle \hat
S_{k,1}^z + \hat S_{k,2}^z \rangle/2$ can be obtained by the
differentiation of the statistical sum (\ref{Z1}) after applying the
decoration-iteration transformation \cite{lis1}:
\[
m_{\textrm{H}} =
\frac{1}{2\beta} \left ( \frac{1}{A}\frac{\partial A}{\partial h_{\textrm{H}}}
+ q_{\textrm{I}\textrm{I}}(1) \frac{\partial (\beta R)}{\partial h_{\textrm{H}}}
+ m_{\textrm{I}} \frac{\partial (\beta h_0)}{\partial h_{\textrm{H}}} \right ).
\]
From $m_{\textrm{I}}$ and $m_{\textrm{H}}$, the summary magnetization can be determined as follows:
\[
m=(m_{\textrm{I}} + 2m_{\textrm{H}})/3.
\]
Now, let us calculate the susceptibility to the action of the
magnetic field $h=h_{\textrm{I}}=h_{\textrm{H}}/r_g$, where $r_g$ is
the ratio of the $g$-factor of a Heisenberg spin to the $g$-factor
of an Ising one:
\[
\chi=\frac{\partial m}{\partial h} = \frac{1}{3}
\left(\frac{\partial m_{\textrm{I}}}{\partial h_{\textrm{I}}} +
\frac{\partial m_{\textrm{I}}}{\partial h_{\textrm{H}}} r_g \right)
+ \frac{2}{3} \left(\frac{\partial m_{\textrm{H}}}{\partial
h_{\textrm{I}}} + \frac{\partial m_{\textrm{H}}}{\partial
h_{\textrm{H}}} r_g \right).
\]

In the particular cases $I_1=I_2$ and $I_2=0$ (or $I_1=0$), the
obtained results agree with the available data for diamond
\cite{dos2} and simple \cite{dos1,lis1} Ising--Heisenberg chains,
respectively.

\section{Numerical Results and Discussion}

Consider the properties of a system with the antiferromagnetic Ising
interaction ($I_1, I_2>0$) and the antiferromagnetic $XXZ$ Heisenberg
interaction: $J_{1}=J_{2}=J\Delta$, $J_{3}=J$, where $\Delta$ is the
interaction anisotropy parameter, $J>0$. In this case, the system is
geometrically frustrated. The magnetic fields $h_{\textrm{I}}$ and
$h_{\textrm{H}}$ are supposed to be equal:
$h=h_{\textrm{I}}=h_{\textrm{H}}$, i.e. $r_g=1$. Without loss of
generality, we assume that $I_1 \geq I_2$ and introduce the
difference of the Ising interaction parameters $\Delta I = I_1 -
I_2$ \cite{dos3}. Passing to the dimensionless parameters, we have
\[
\tilde{J}=\frac{J}{I_1},  \quad \Delta \tilde{I}=\frac{\Delta I}{I_1}, \quad \tilde{h}=\frac{h}{I_1}.
\]
The parameter $\Delta \tilde{I}$ has a physical sense on the
interval $[0,1]$ and characterizes the degree of asymmetry of the
Ising interaction at the bonds along the diamond sides.

First, we consider the properties of the ground state of the system.
It corresponds to the lowest energy $\tilde{\cal E}_i={\cal E}_i/I_1$
of spectrum (\ref{Ek}) for the possible configurations
$\hat\mu^z_{k}$ and $\hat\mu^z_{k+1}$. Depending on the parameters
$\tilde{J}$, $\Delta$, $\Delta \tilde{I}$, and $\tilde{h}$, the
system can have four ground states: saturated paramagnetic state
(SPA), ferrimagnetic state (FRI), unsaturated paramagnetic state
(UPA), and nodal antiferromagnetic state (NAF). The energies of
these states for the initial cell are as follows:
\[
\tilde{\cal E}_{\textrm{SPA}} = \frac{\tilde{J}}{4} + 1 - \frac{\Delta \tilde{I}}{2} - \frac{3\tilde{h}}{2},
\]
\[
\tilde{\cal E}_{\textrm{FRI}} = \frac{\tilde{J}}{4} - 1 + \frac{\Delta \tilde{I}}{2} - \frac{\tilde{h}}{2},
\]
\[
\tilde{\cal E}_{\textrm{UPA}} = - \frac{\tilde{J}}{4} - \frac{\tilde{J}\Delta}{2} - \frac{\tilde{h}}{2},
\]
\[
\tilde{\cal E}_{\textrm{NAF}} = - \frac{\tilde{J}}{4}- \frac{1}{2}\sqrt{\tilde{J}^{2} \Delta^2 + \Delta \tilde{I}^2}.
\]
The wave functions of these states have the form
\[
|\mbox{SPA} \rangle = \prod\limits_{k=1}^N| + \rangle_k ~|\uparrow, \uparrow \rangle_{k,1;k,2},
\]
\[
|\mbox{FRI} \rangle = \prod\limits_{k=1}^N| - \rangle_k ~|\uparrow, \uparrow \rangle_{k,1;k,2},
\]
\[
|\mbox{UPA} \rangle {=} \left\{\!\! \begin{array}{ll}
\prod\limits_{k=1}^N \! |{+}\rangle_k\!
\left[\frac{1}{\sqrt{2}}(|\uparrow, \downarrow \rangle {-} |\downarrow, \uparrow \rangle) \right]_{k,1;k,2}
&\! \textrm{for}~ \Delta{\neq}0, \\[4mm]
\prod\limits_{k=1}^N \! |{+}\rangle_k\! \left|\uparrow, \downarrow
\atop \downarrow, \uparrow \right\rangle_{k,1;k,2} &\! \textrm{for}~
\Delta{=}0,
\end{array} \right.
\]
\[
|\mbox{NAF} \rangle {=} \prod\limits_{k=1}^N \left|(-)^{n={\left\{{k~~}\atop{k+1}\right.}} \! \right\rangle_k \times
\]
\[
\times \left[A^{(-)^n} |\uparrow, \downarrow \rangle - A^{(-)^{n{+}1}} |\downarrow, \uparrow \rangle \right]_{k,1;k,2},
\]

%
\begin{figure}
\includegraphics[width=8.2cm]{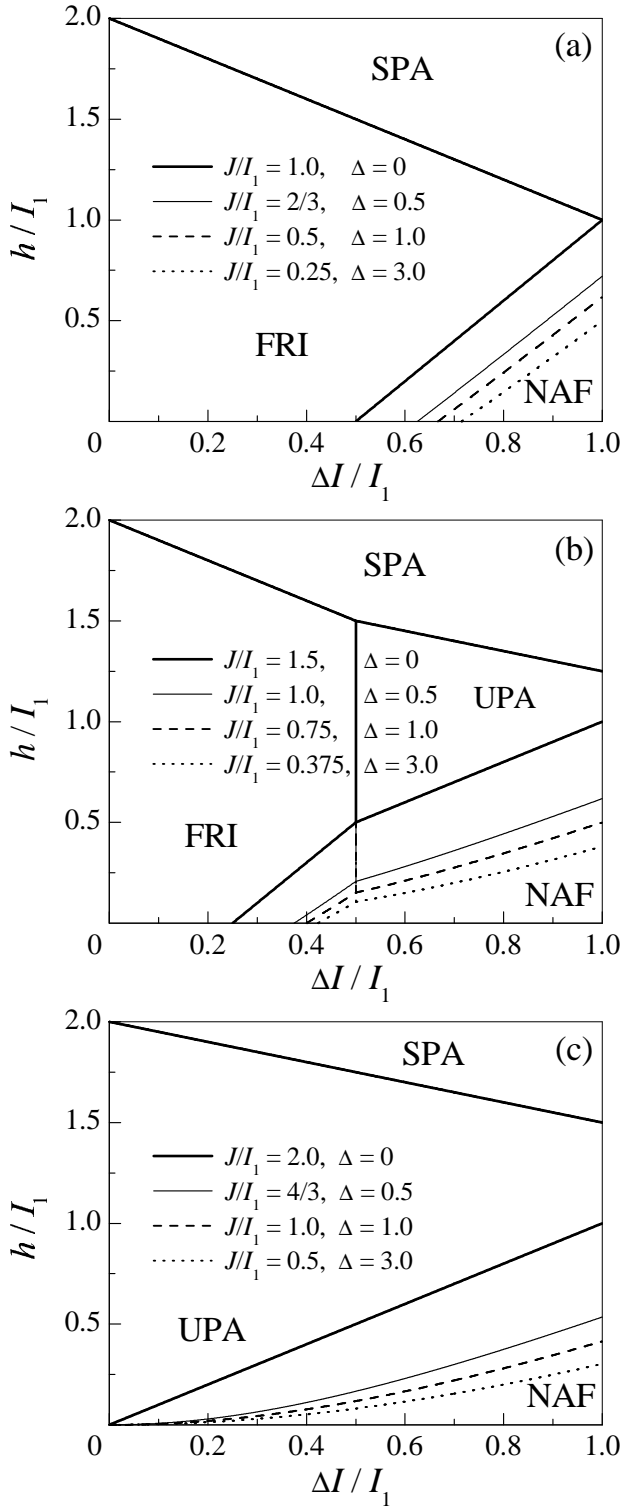}
\caption{Ground-state phase diagram $(\Delta\tilde{I},\tilde{h})$ as
a function of $\tilde{J}$ and $\Delta$. Three characteristic
topologies are shown: ({\it a}), ({\it b}), and ({\it c}) with
transition lines for several sets of values of $\tilde{J}$ and
$\Delta$  }\label{fig2}
\end{figure}

\noindent where the vectors $| \pm \rangle_k$ describe the state of
the spins $\hat \mu^z_k$: $|+\rangle_k=|\uparrow\rangle_k$,
$|-\rangle_k=|\downarrow\rangle_k$. The wave functions of the doubly
degenerated NAF state are written down with the help of the
expression $(-)^n \in \{-,+\}$ that means the sign of the number
$(-1)^n$. The coefficients $A^{\pm}$ are as follows:
\[
A^{\pm}=\frac{1}{\sqrt{2}}\sqrt{1 \mp \frac{\Delta \tilde{I}}{\sqrt{\tilde{J}^{2} \Delta^2 + \Delta \tilde{I}^2}}}.
\]
In the UPA state at $\Delta=0$, the pair of spins $\hat {\bf
S}_{k,1}$ and $\hat {\bf S}_{k,2}$ occupies one of the two
equiprobable  states $|\uparrow, \downarrow \rangle_{k,1;k,2}$ and
$|\downarrow, \uparrow \rangle_{k,1;k,2}$. That is why this state is
macroscopically degenerated and has the residual entropy
$s_{\textrm{res}}=k_{\textrm{B}} \ln2$.

The ground-state phase diagram $(\Delta \tilde{I},\tilde{h})$ can
have topology of three types depending on $\tilde{J}(1+\Delta)$
(Fig.~\ref{fig2}). The phase diagram with topology of type 1
(Fig.~\ref{fig2},{\it a}) includes the FRI, NAF, and SPA states. It
is realized at $\tilde{J}(1+\Delta)\leq1$. In the zero field, the FRI
and NAF states are separated by the critical point
\[
\Delta \tilde{I}_{\textrm{F.N}} = \frac{2-\tilde{J}}{2} - \frac{\tilde{J}^2 \Delta^2}{2(2-\tilde{J})}.
\]
It is worth noting that, in the FRI state, there arises the
geometrical frustration effect on the Heisenberg bond: the pair of
Heisenberg spins occupies the ferromagnetic state $|\uparrow,
\uparrow \rangle_{k,1;k,2}$ that does not correspond to the minimum
of the energy $J\hat S^z_{k,1}\hat S^z_{k,2}$. In the NAF state, one
observes the effect of quantum fluctuations on this bond: the pair
of Heisenberg spins occupies the mixed states $\left[A^{\mp}
|\uparrow, \downarrow \rangle - A^{\pm} |\downarrow, \uparrow
\rangle \right]_{k,1;k,2}$ with the minimum energy $J\hat
S^z_{k,1}\hat S^z_{k,2}$. The phase diagram with topology of
type 2 (Fig.~\ref{fig2},{\it b}) includes all the ground states and
is realized at $1< \tilde{J}(1 + \Delta) <2$. The FRI and UPA states
are separated by the line $\Delta\tilde{I}=
\Delta\tilde{I}_{\textrm{F|U}}$, where
$\Delta\tilde{I}_{\textrm{F|U}} = 2-\tilde{J}(1+\Delta)$. The phase
diagram with topology of type 3 (Fig.~\ref{fig2},{\it c}) includes
the NAF, UPA, and SPA states. It is realized at
$2\leq\tilde{J}(1+\Delta)$. The line of transition between the NAF
and UPA states starts at the point $(0,0)$. In Fig.~\ref{fig2}, one
can see that the interval $(0,\Delta \tilde{I}_{\textrm{F.N}})$, in
which there arises the geometrical frustration effect on the
Heisenberg bond, decreases to zero due to the intensification of
quantum fluctuations.

The SPA and FRI states are identical, whereas the UPA (except for
the case of $\Delta=0$) and NAF states are isomorphous to the
corresponding states of an asymmetric diamond Ising--Hubbard chain
\cite{dos3,lis2}. The typical topologies of the ground-state phase
diagram $(\Delta \tilde{I},\tilde{h})$ are also the same as those for this
chain \cite{dos3,lis2}. That is why the specific features of the
ground state at the critical points of the phase diagrams $(\Delta
\tilde{I},\tilde{h})$ described in \cite{lis2} are also observed at
the corresponding points in Fig.~\ref{fig2} at $\Delta\neq0$. At
$\Delta=0$, the ground-state characteristics at the critical points,
where the UPA state is realized, can have values other than those at
$\Delta\neq0$. In particular, at the critical point $(0,0)$ in
Fig.~\ref{fig2},{\it c}, the residual entropy at $\Delta=0$ is
larger than that at $\Delta\neq0$, namely
$s_{\textrm{res}}=k_{\textrm{B}} \ln5$ for $\tilde{J}=2$ and
$s_{\textrm{res}}=k_{\textrm{B}} \ln4$ for $\tilde{J}>2$, where the
ground state is frustrated \cite{lis2}. In addition, the ground
state of our system has interesting peculiarities at $\tilde{J}=1$
and $\Delta=0$ at the critical point $(1,1)$ in Fig.~\ref{fig2},{\it
a}. It is the only point of coexistence of all the ground states
(SPA, FRI, NAF, and UPA). In addition, it is the point, at which the
nodal antiferromagnetic state NAF$_{+}$ is realized with the energy
$\tilde{\cal E}_{\textrm{NAF}_+}={\tilde{J}}/{4}-\tilde{h}$ and the
wave function
\[
|\mbox{NAF}_{+} \rangle = \prod\limits_{k=1}^N \left|(-)^{n=\left\{{k~~}\atop{k+1}\right.} \!\!\right\rangle_k
|\uparrow, \uparrow \rangle_{k,1;k,2}.
\]
The Ising subsystem has the characteristics
\[
\beta R {=} \ln\frac{3}{4}, ~\beta h_0 {=}\ln 3, ~m_{\textrm{I}}{=}\frac{m_{\textrm{s}}}{\sqrt{5}},
~q_{\textrm{I}\textrm{I}}(n){=}\frac{1 + 4(4\sqrt{5}{-}9)^n}{20},
\]
where $m_{\textrm{s}}=1/2$ is the saturation magnetization. In this
ground state, $R=h_0=0$, but Ising spins are not effectively free,
because the temperature dependence of $R$ and $h_0$ has a linear
component \cite{lis2}. This ground state has $s_{\textrm{res}} =
k_{\textrm{B}} \ln (2+\sqrt{5})$.

The ground-state phase diagrams ($\Delta\tilde{I},\tilde{h}$) in
Fig.~\ref{fig2} demonstrate the following regularities. The diamond
Ising--Heisenberg chain can have the NAF ground state if
$\Delta\tilde{I}\neq0$ in the chain. The simple diamond
Ising--Heisenberg chain can have the FRI ground state in the zero
field, if the interaction of an Ising spin with second neighbors is
strong enough ($\Delta\tilde{I}<1$).

\begin{figure}
\includegraphics[width=\column]{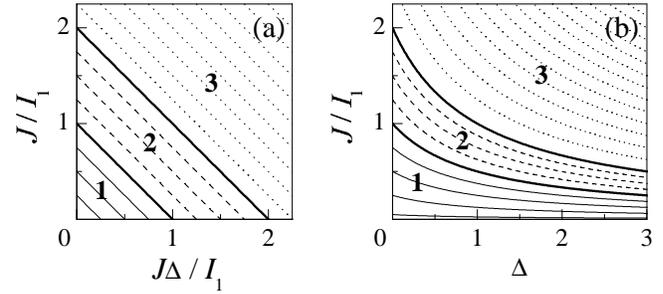}
\vskip-3mm\caption{Topological diagrams
($\tilde{J}\Delta,\tilde{J}$) ({\it a}) and ($\Delta,\tilde{J}$)
({\it b}) of the ground-state phase diagram
$(\Delta\tilde{I},\tilde{h})$ covered by ``equitopological'' lines
(\ref{JD}). Thick lines mark the boundary between the regions of
three typical topologies specified by the corresponding numbers
}\label{fig3}
\end{figure}

Now, we consider the influence of the Heisenberg interaction on the
ground-state phase diagram ($\Delta\tilde{I},\tilde{h}$). In the
case if the energy $\tilde{\cal E}_{\textrm{UPA}}$ remains constant
with respect to the energy $\tilde{\cal E}_{\textrm{FRI}}$ (which
means that it also remains constant with respect to $\tilde{\cal
E}_{\textrm{SPA}}$), a change of $\tilde{J}$ and $\Delta$ gives rise
to a much simpler reconstruction of the phase diagram
($\Delta\tilde{I},\tilde{h}$) than their independent variation. Let
us determine the constant of this condition at the point
($\tilde{J}=\tilde{J}^*,\Delta=0$), where the Heisenberg interaction
is converted to the Ising one $\tilde{J}^*$. As a result, we obtain the
interrelation of $\tilde{J}$ and $\Delta$ as follows:
\begin{equation}
\tilde{J} + \tilde{J}\Delta = \tilde{J}^*.
\label{JD}
\end{equation}
According to the above-described dependence of the topology of the
phase diagram ($\Delta\tilde{I},\tilde{h}$) on the Heisenberg
interaction, the topologically equivalent phase diagrams
($\Delta\tilde{I},\tilde{h}$) are obtained in mode (\ref{JD}). The
change of the phase diagram ($\Delta\tilde{I},\tilde{h}$) in mode
(\ref{JD}) is shown in Fig.~\ref{fig2}, where the sets of values
$\tilde{J}$ and $\Delta$ for each case ({\it a, b}, or {\it c})
correspond to a certain $\tilde{J}^*$. One can see that this change
consists in a shift of the lines of the NAF~$\leftrightarrow$~FRI
and NAF~$\leftrightarrow$~UPA transitions. Based on the above consideration,
a diagram reflecting the influence of the Heisenberg interaction on
the topology of the phase diagram ($\Delta\tilde{I},\tilde{h}$) can
be constructed. Such a topological diagram is presented in
Fig.~\ref{fig3} in two planes: ($\tilde{J}\Delta,\tilde{J}$) and
($\Delta,\tilde{J}$). The topological diagram
($\tilde{J}\Delta,\tilde{J}$) illustrates the fact that the topology
of the phase diagram ($\Delta\tilde{I},\tilde{h}$) depends on the
$ZZ$- and $XY$-components of the Heisenberg interaction in the same
way.

\begin{figure*}
\includegraphics[width=17.5cm]{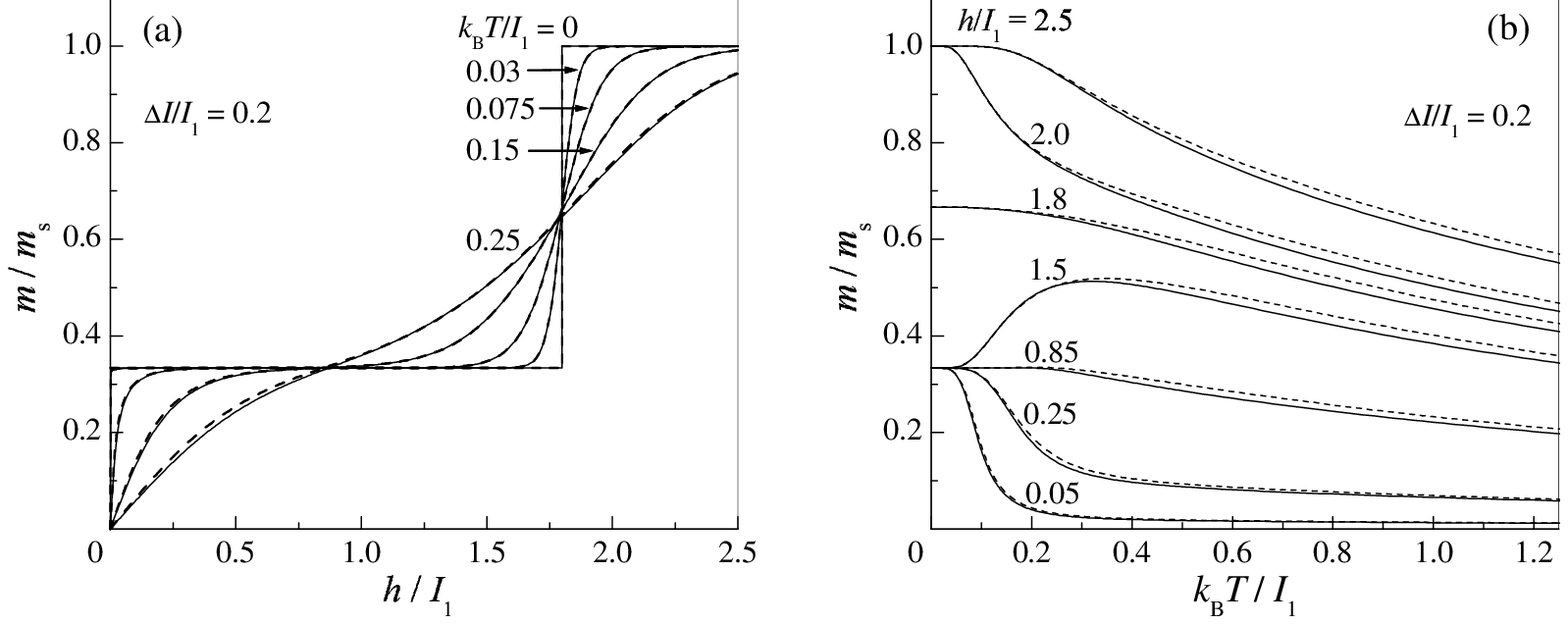}
\vskip-4mm\caption{Summary magnetization as a function of the field
for different temperatures ({\it a}) and as a function of the
temperature for different fields ({\it b}) in the case of the FRI
ground state in the zero field. Solid lines mark the results
obtained for $\tilde{J}=1.0$ and $\Delta=0.5$, dashed lines~-- those
obtained for $\tilde{J}=0.375$ and $\Delta=3$ }\vskip7mm\label{fig4}
\end{figure*}

\begin{figure*}
\includegraphics[width=17.5cm]{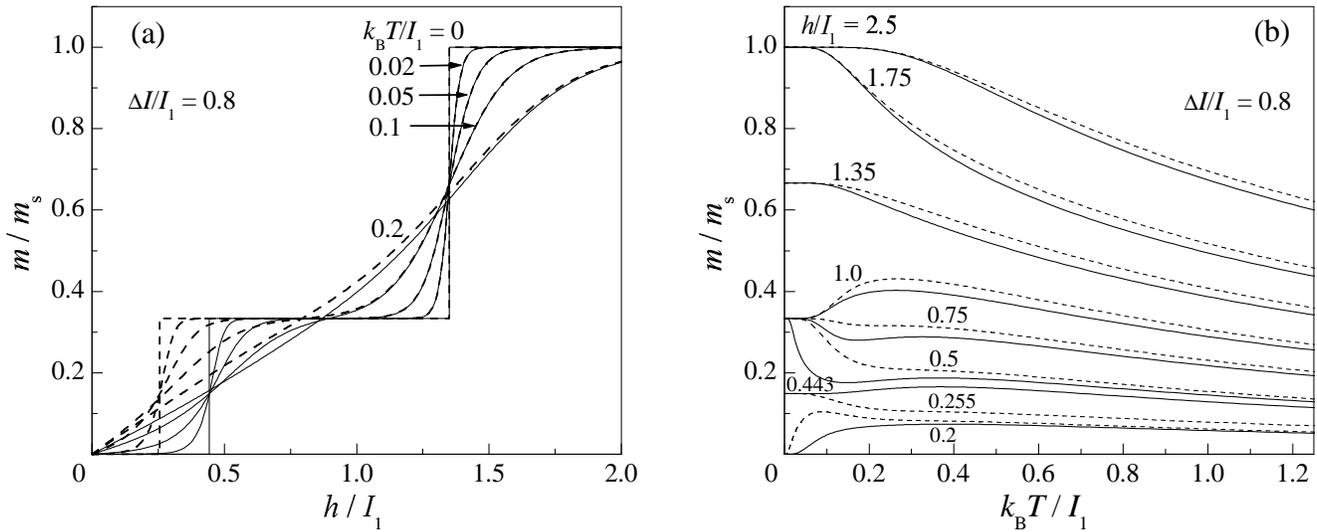}
\vskip-4mm\caption{Summary magnetization as a function of the field
for different temperatures ({\it a}) and as a function of the
temperature for different fields ({\it b}) in the case of the NAF
ground state in the zero field. Solid lines mark the results
obtained for $\tilde{J}=1.0$ and $\Delta=0.5$, dashed lines~-- those
obtained for $\tilde{J}=0.375$ and $\Delta=3$  }\label{fig5}
\end{figure*}

Let us consider the magnetization process and thermodynamic
characteristics as functions of the temperature and their dependence
on the Ising interaction asymmetry and Heisenberg interaction
parameters in mode (\ref{JD}). For this purpose, we select sets of
values of $\tilde{J}$ and $\Delta$, the ground-state phase diagram
($\Delta\tilde{I},\tilde{h}$) for which is depicted in
Fig.~\ref{fig2},{\it b}. The summary magnetization as a function of
the field for different temperatures and as a function of the
temperature for different fields is presented for two characteristic
cases: the FRI ground state in the zero field (Fig.~\ref{fig4}) and
the NAF ground state in the zero field (Fig.~\ref{fig5}). The field
dependence of the summary magnetization at the zero temperature in
Fig.~\ref{fig4},{\it a} has an intermediate plateau
$m/m_{\textrm{s}}=1/3$ corresponding to the FRI ground state,
whereas in Fig.~\ref{fig5},{\it a}, there are two intermediate
plateaus: $m/m_{\textrm{s}}=0$ corresponding to the NAF ground state
and $m/m_{\textrm{s}}=1/3$ corresponding to the UPA ground state.
The magnetization $m/m_{\textrm{s}}$ in Fig.~\ref{fig4},{\it b} in
the critical field of the FRI~$\leftrightarrow$~SPA transition tends
to $2/3$ as the temperature approaches zero. In Fig.~\ref{fig5},{\it
b}, the magnetization $m/m_{\textrm{s}}$ in the lower and upper
critical fields corresponding to the NAF~$\leftrightarrow$~UPA  and
UPA~$\leftrightarrow$~SPA transitions tends to $1/(3\sqrt{5})$ and
$2/3$, respectively. The Heisenberg interaction significantly
influences the field dependence of $m$ at low temperatures in the
field range belonging to the region of the NAF ground state and a
certain part of the neighboring region (Fig.~\ref{fig5},{\it a}). As
a result, it also has some effect on the low-temperature dependence
of $m$ for fields belonging to this region (Fig.~\ref{fig5},{\it
b}). Due to the intensification of quantum fluctuations, the
temperature curve of $m$ shifts upward in the region of medium and
high temperatures (Fig.~\ref{fig4},{\it b} and Fig.~\ref{fig5},{\it
b}). This increment of $m$ is a result of the decrease of
$m_{\textrm{I}}$ due to a reduction of the effective field $h_0,$ as
well as the increase of $m_{\textrm{H}}$ due to the relative growth
of the energies $\hat {\cal H}_k$ (\ref{Ek}) corresponding to states
with zero magnetization of the Heisenberg spins.

\begin{figure}
\includegraphics[width=\column]{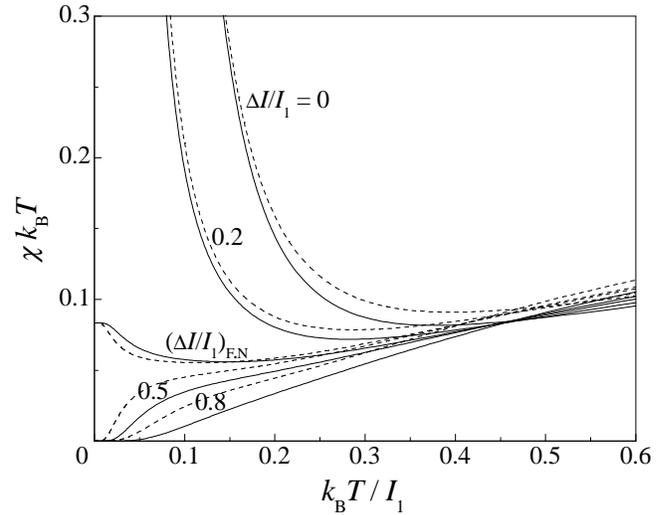}
\caption{Magnetic susceptibility multiplied by the temperature as a
function of the temperature in the zero field. Solid lines mark the
results obtained for $\tilde{J}=1.0$ and $\Delta=0.5$, dashed lines
~-- those obtained for $\tilde{J}=0.375$ and $\Delta=3$
}\label{fig6}
\end{figure}

The magnetic susceptibility multiplied by the temperature ($\chi
k_{\textrm{B}} T$) as a function of the temperature in the zero field is
presented in Fig.~\ref{fig6}. With the FRI ground state in the zero
field, this dependence looks like that for quantum ferrimagnetics
\cite{yam99}, while, in the case of the NAF ground state in the zero
field, it has an antiferromagnetic character. As the temperature
tends to zero, the quantity $\chi k_{\textrm{B}} T$ can either
exponentially diverge (if $\Delta \tilde{I}$ corresponds to the FRI
ground state), or exponentially tends to zero (if $\Delta \tilde{I}$
corresponds to the NAF ground state), or takes the value of 1/12
(if $\Delta \tilde{I}$ appears exactly in the critical point $\Delta
\tilde{I}_{\textrm{F.N}}$). The low-temperature dependence of $\chi
k_{\textrm{B}} T$ strongly reacts to a variation of the quantum
fluctuation intensity, if $\Delta \tilde{I}$ appears in some
neighborhood of the critical point $\Delta \tilde{I}_{\textrm{F.N}}$
or above it (Fig.~\ref{fig6}). The high-temperature dependence of
$\chi k_{\textrm{B}} T$ shifts to a higher susceptibility with
increase in the intensity of quantum fluctuations.

\begin{figure*}
\includegraphics[width=17.5cm]{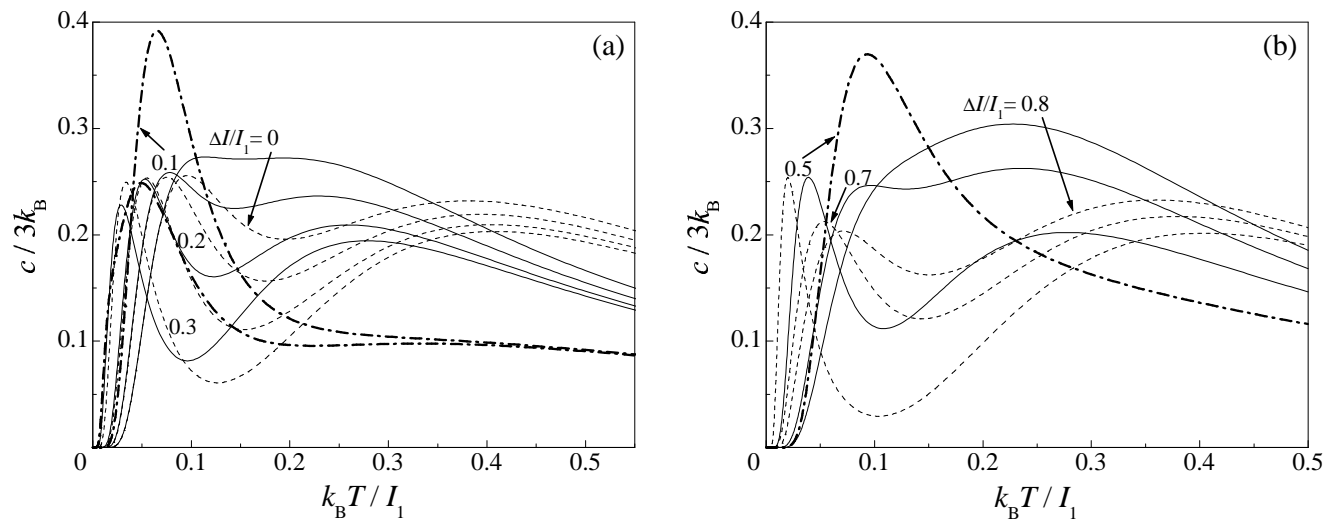}
\caption{Temperature dependence of the heat capacity in the zero field
at $\Delta \tilde{I}<\Delta \tilde{I}_{\textrm{F.N}}$ ({\it a}) and
$\Delta \tilde{I}>\Delta \tilde{I}_{\textrm{F.N}}$ ({\it b}).
Dash-and-dot lines mark the results obtained for $\tilde{J}=1.5$ and
$\Delta=0$, solid lines -- for $\tilde{J}=1.0$ and $\Delta=0.5$, and
dashed lines -- for $\tilde{J}=0.375$ and $\Delta=3$  }\label{fig7}
\end{figure*}

For a certain interval of $\Delta \tilde{I}$ widening with increase
in the quantum fluctuation intensity, the temperature dependence of
the heat capacity in the zero field has two maxima: principal and
low-temperature ones (Fig.~\ref{fig7}). At $\Delta=0$, it
particularly has two maxima for $\Delta \tilde{I}\in(0.16,0.31)$,
and one intense low-temperature maximum beyond this interval that
considerably changes with the appearance of weak quantum
fluctuations (Fig.~\ref{fig7}). An increase of their intensity
results in a considerable growth of the height of the principal
maximum and a decrease of its temperature. With their further
intensification, however, the height of the principal maximum
decreases, while its temperature rises (Fig.~\ref{fig7}). The growth
of $\Delta \tilde{I}$ in the interval $(0,\Delta
\tilde{I}_{\textrm{F.N}})$ results in a reduction of the height of
the principal maximum and the rise of its temperature, whereas, in
the interval $(\Delta \tilde{I}_{\textrm{F.N}},1)$, the situation is
opposite. The low-temperature maximum is mainly caused by thermal
excitations responsible for the transitions between the FRI, NAF,
and UPА states. Its height and temperature considerably change due
to the variation of the intensity of quantum fluctuations, if
$\Delta \tilde{I}$ lies in a certain vicinity of the critical point
$\Delta \tilde{I}_{\textrm{F.N}}$ or above it (Fig.~\ref{fig7}).
With increase in $\Delta \tilde{I}$, its temperature noticeably
falls for $\Delta \tilde{I}<\Delta \tilde{I}_{\textrm{F.N}}$ and
grows for $\Delta \tilde{I}>\Delta \tilde{I}_{\textrm{F.N}}$. If
$\Delta \tilde{I}$ appears in a rather close vicinity of the
critical point $\Delta \tilde{I}_{\textrm{F.N}}$, the thermal
excitation corresponding to the energy of the
FRI~$\leftrightarrow$~NAF transition gives rise to the formation of
an additional low-temperature maximum close to the zero temperature.
The formation and evolution of this maximum depending on $\Delta
\tilde{I}$ takes place similarly to the heat capacity of an
asymmetric diamond Ising--Hubbard chain~\cite{lis2}.

\section{Conclusions}

The ground state and the thermodynamics of a spin-1/2 asymmetric
diamond Ising--Heisenberg chain are investigated. Exact calculations
of the free energy, entropy, heat capacity, magnetization of Ising
and Heisenberg spins, and magnetic susceptibility are performed by
the method of decoration-iteration transformation for the $XYZ$
anisotropic Heisenberg interaction. In the case of the
antiferromagnetic Ising interaction and antiferromagnetic $XXZ$
Heisenberg interaction, with the system being geometrically
frustrated, the ground state, field and temperature dependences of
the magnetization, and temperature dependence of the magnetic
susceptibility and of the heat capacity in zero field are
investigated. The influence of the asymmetry parameter of the Ising
interaction ($\Delta \tilde{I}$) and the Heisenberg interaction
parameters ($\tilde{J}$ and $\Delta$) on these characteristics in
the mode $\tilde{J}+\tilde{J}\Delta = \rm{const}$ (\ref{JD}) is
studied.

The geometrically frustrated system under study has four ground
states: SPA, FRI, UPА, and NAF. Its ground state phase diagram
($\Delta\tilde{I},\tilde{h}$)  has three typical topologies, which
is similar to the situation observed in the case of an asymmetric
diamond Ising--Hubbard chain \cite{dos3,lis2}. A variation of the
Heisenberg interaction parameters in mode (\ref{JD}) does not change
the topology of the phase diagram ($\Delta\tilde{I},\tilde{h}$); it
only shifts the boundaries of the NAF ground state at the latter.
The topology of the phase diagram ($\Delta\tilde{I},\tilde{h}$)
depending on the Heisenberg interaction is described by the
topological diagrams ($\tilde{J}\Delta,\tilde{J}$) and
($\Delta,\tilde{J}$). It is shown that, at the zero temperature, rather
strong quantum fluctuations remove the geometrical frustration
effect on the Heisenberg bond.

The intensification of quantum fluctuations results in an increase
of the summary magnetization and the magnetization of Heisenberg spins and
in a decrease of the magnetization of Ising spins in the region of
medium and high temperatures. In this case, a growth of the magnetic
susceptibility in the zero field is observed. The temperature dependence
of the heat capacity in the zero field has a principal maximum and a
low-temperature one for a certain interval of $\Delta \tilde{I}$
widening with increase in the quantum fluctuation intensity. The
heights and the temperatures of these maxima can significantly change
depending on the strength of quantum fluctuations and noticeably
change depending on $\Delta \tilde{I}$. If $\Delta \tilde{I}$ lies
in a rather close vicinity of the critical point $\Delta
\tilde{I}_{\textrm{F.N}}$, the heat capacity has another
low-temperature maximum caused by the transition between the FRI and
NAF states.

The obtained results are also valid for a simple Ising--Heisenberg
chain, in which an Ising spin interacts with the first and second
neighbors. In the limiting cases $I_1=I_2$ ($\Delta \tilde{I}=0$)
and $I_2=0$ ($\Delta \tilde{I}=1$), these results correspond to the
earlier considered diamond \cite{dos2} and simple \cite{dos1,lis1}
Ising--Heisenberg chains.

\vskip3mm The author is grateful to Prof. O.V.~Derzhko and
Dr.~Т.М.~Verkholyak for the discussion and useful remarks.

\rezume{%
СПІН-1/2 АСИМЕТРИЧНИЙ РОМБІЧНИЙ ЛАНЦЮЖОК\\
ІЗИНГА--ГАЙЗЕНБЕРГА}{Б.М.~Лісний} {Розглянуто основний стан і
термодинаміку спін-1/2 асиметричного ромбічного ланцюжка
Ізинга--Гайзенберга. Для $XYZ$ анізотропної взаємодії Гайзенберга
методом декораційно-іте\-ра\-цій\-но\-го перетворення точно
розраховано вільну енергію,\rule{0pt}{10pt} ентропію, теплоємність,
намагніченість і магнітну сприйнятливість.\rule{0pt}{10pt} У
випадку\rule{0pt}{10pt} антиферомагнітних взаємодій -- Ізинга і
$XXZ$ анізотропної\rule{0pt}{10pt} Гайзенберга -- досліджено
основний стан, процес намагнічування,\rule{0pt}{10pt} температурну
залежність намагніченості, магнітної\rule{0pt}{10pt} сприйнятливості
і теплоємності. Вивчено вплив геометричної\rule{0pt}{10pt}
фрустрації та квантових флуктуацій на ці
характеристики.\rule{0pt}{10pt}}

\end{document}